\documentclass[twocolumn,showpacs,amsmath,amssymb,nofootinbib]{revtex4}
\usepackage{graphicx}
\usepackage{dcolumn}
\usepackage{bm}

\def\simge{\mathrel{
     \rlap{\raise 0.511ex \hbox{$>$}}{\lower 0.511ex \hbox{$\sim$}}}}
\def\simle{\mathrel{
     \rlap{\raise 0.511ex \hbox{$<$}}{\lower 0.511ex \hbox{$\sim$}}}}
\def\be{\begin{equation}}
\def\ee{\end{equation}}
\def\bea{\begin{eqnarray}}
\def\eea{\end{eqnarray}}
\def\dblone{\hbox{$1\hskip -1.2pt\vrule depth 0pt height 1.6ex width 0.7pt
                  \vrule depth 0pt height 0.3pt width 0.12em$}}

\newcommand{\boldl}{{\bf L}}
\newcommand{\boldu}{{\bf U}}
\newcommand{\tr}{{\rm tr}\,}
\renewcommand{\d}{{\rm d}}

\begin{document}
\title{Effective potential for Polyakov loops from a center
  symmetric effective theory in three dimensions}

\author{Dominik Smith$^{a,b,c}$}
\affiliation{
$^a$Institut f\"ur Theoretische Physik,
Johann Wolfgang Goethe-Universit\"at,
Max-von-Laue-Str.\ 1, 60438 Frankfurt am Main, Germany\\
$^b$Frankfurt International Graduate School for Science (FIGGS),
Ruth-Moufang-Str.\ 1, 60438 Frankfurt am Main, Germany\\
$^c$Helmholtz Research School for Quark Matter Studies (H-QM),
Max-von-Laue-Str.\ 1, 60438 Frankfurt/Main, Germany
}

\date{\today}
\begin{abstract}
We present lattice simulations of a center symmetric dimensionally
reduced effective field theory for SU(2) Yang Mills which employ
thermal Wilson lines and three-dimensional magnetic fields as
fundamental degrees of freedom. The action is composed of a gauge
invariant kinetic term, spatial gauge fields and a potential for the
Wilson line which includes a "fuzzy" bag term to generate
non-perturbative fluctuations. The effective potential for the
Polyakov loop is extracted from the simulations including all modes of
the loop as well as for cooled configurations where the hard modes have
been averaged out. The former is found to exhibit a non-analytic
contribution while the latter can be described by a mean-field like
{\em ansatz} with quadratic and quartic terms, plus a Vandermonde
potential which depends upon the location within the phase diagram.
\end{abstract}
\maketitle

\section{Introduction} \label{sec:Intro}

QCD at temperatures $T\simeq 200$~MeV$-1$~GeV exhibits only partial
deconfinement and, moreover, perturbation theory fails to reproduce
some thermodynamic quantities such as the equation of state
\cite{Andersen:2004fp}. Several authors
\cite{dlp,renpol,Meisinger,Pisarski:2006hz,Vuorinen:2006nz}
have suggested that in this regime a more appropriate effective
description is not in terms of quasi-particles but in terms of the
thermal Wilson line
\bea
\boldl(\bm{x}) &=& {\cal Z}_R^{-1} \,
    {\cal P} \exp\left( ig\int\limits_0^{1/T} \d\tau\;
      A_0(\bm{x},\tau) \right)~, \label{WilsonL}
\eea
and the spatial components of the gauge field.  The operator $A_0$
from eq.~(\ref{WilsonL}) is defined on four dimensional euclidean
space-time while $\boldl(\bm{x})$ is a matrix-valued field in
space. ${\cal Z}_R$ denotes a renormalization constant, which can
be calculated in different representations of $\boldl$ 
\cite{renpol,Kaczmarek:2002mc}. We take $\boldl$ to be
in the fundamental representation here.
In such a
framework, the deconfined phase is not a free gas, but rather a
condensate of spin-like operators 
\bea \ell(\bm{x}) &=& \frac{1}{N}
\tr\boldl(\bm{x})~, \label{RenPL} 
\eea
called \emph{Polyakov loops}. The volume averaged expectation value
of this operator is an order parameter for the deconfining phase
transition in the limit of infinitely massive quarks
\cite{Kaczmarek:2002mc}. In contrast to ferromagnetism, the high
temperature phase is the ordered phase where a global symmetry is
spontaneously broken.

We perform lattice simulations of an effective theory in three
dimensions defined in the continuum by~\cite{Pisarski:2006hz}
\bea 
{\cal L}^{\rm eff} &=& \frac{1}{2}\tr G^2_{ij} + \frac{T^2}{g^2}
   \tr |\boldl^\dagger D_i \boldl|^2 \nonumber \\ 
& & - \frac{2}{\pi^2} T^4
  \sum\limits_{n\ge1} \frac{1}{n^4} |\tr \boldl^n|^2 + B_f T^2 |\tr\boldl|^2~.
	\label{Leff_cl} 
\eea
All fields in~(\ref{Leff_cl}) are functions of $\bm{x}$ only. $G_{ij}$
is the magnetic field strength. The second term is the contribution
from electric fields since in the three-dimensional theory, for {\em
  arbitrary} $A_0$, ${\bf E}$ is given by~\cite{Pisarski:2006hz}
\be
{\bf E}_i(\bm{x}) = \frac{T}{ig} \, \boldl^\dagger (\bm{x}) 
 {\bf D}_i(\bm{x}) \boldl(\bm{x}) ~.
\ee
The potential $\sim \, -\sum_{n\ge1} |\tr \boldl^n|^2/n^4$ for the
Wilson line $\boldl$ is obtained by computing the one-loop fluctuation
determinant in a constant background ${\bf A}_0$ (or $\boldl$)
field~\cite{GPY}. It is evidently minimized by the
perturbative vacuum $\langle\boldl\rangle=\dblone$ (times a phase),
for any $T$. To generate a phase transition in infinite volume,
refs.~\cite{Meisinger,Pisarski:2006hz} suggested to add non-perturbative
contributions such as $B_f T^2 |\tr\boldl|^2$, with $B_f$ a ``fuzzy''
bag constant (see, also, refs.~\cite{Megias}). The ``fuzzy
bag'' dominates at sufficiently low temperature and induces a
transition to a confined phase with $\langle\tr\boldl\rangle=0$.

The three-dimensional effective theory~(\ref{Leff_cl}) is valid
only over distance scales larger than $1/T$ and is non-renormalizable.
A related renormalizable 3D theory, which in a rough sense
corresponds to a linear sigma model version of~(\ref{Leff_cl}) has
been formulated in refs.~\cite{Vuorinen:2006nz} 
(see also ref.~\cite{Altes:2008ic}). These effective theories respect the global
Z(N) center symmetry of the four-dimensional Euclidean SU(N)
Yang-Mills theory. This allows for non-perturbatively large
fluctuations of the Wilson line and it is interesting to investigate
whether such an extension of high-temperature perturbation theory is
sufficient to describe the properties of hot Yang-Mills even close to
the temperature for deconfinement.

We note that there have been various attempts at extracting through
numerical simulations a three-dimensional effective action which
reproduces the long-range properties of the underlying
four-dimensional Yang-Mills theory~\cite{heinzl}. Here, our approach
is different (mostly because we do not aim at matching the couplings
of the 3d theory yet). We shall focus on extracting an effective
potential and analysing its structure from the 3d theory itself.

\section{Lattice action}

Our present simulations have been performed for gauge group SU(2).
The structure of this group is simpler than that of SU(3), and thus allows
for much higher numerical precision but exhibits the qualitative
features which we are interested in, namely a deconfining
phase transition and non-perturbative fluctuations between
center-symmetric states. The lattice action is of the form
\bea 
S&=& \beta \sum_\Box ( 1-\frac{1}{2} \rm {Re\, tr}~\boldu_\Box )  \nonumber\\
& &- \frac{1}{2}\beta\sum\limits_{\langle ij\rangle} \tr ( \boldl_i
\boldu_{ij}\boldl_j^\dagger \boldu_{ij}^\dagger + {\rm h.c.})  - m^2
\sum\limits_i |\tr\boldl_i|^2~. \nonumber\\
\label{Slatt}
\eea
The first term is the standard Wilson action for the magnetic fields
in three dimensions; the sum runs over all spatial plaquettes. We have
checked that our implementation reproduces the plaquette expectation
values published in ref.~\cite{Hietanen:2006rc}. The second term is a
kinetic term for the Wilson line corresponding to the electric fields
in three dimensions. Here, the sum runs over all links connecting
nearest neighbor sites and the gauge links $U_{ij}$ ensure gauge
invariance. The third term is a mass term for the trace of the Wilson
line which combines the $n=1$ contribution to the one-loop potential
with the non-perturbative "fuzzy" bag contribution. Contributions
from larger $n$ have been dropped\footnote{Terms corresponding to $n>1$
 are suppressed by $1/n^4$. They are most
important when the Polyakov loop is near unity; there, the full sum,
$\sum 1/n^4 = \pi^4/90 = 1.08232$, is about $8.2\%$ larger than the leading
term.}. We have previously
performed simulations of a simplified version of~(\ref{Slatt}) without
magnetic fields in ref.~\cite{DS}. For $m^2=0$, and without magnetic
fields, our code reproduces free energy measurements from refs.\
\cite{Guha:1983ib} but differs slightly from the older work
of ref.\ \cite{KSS} which used smaller lattices and lower statistics.

We employ the standard Metropolis algorithm to generate a thermal
ensemble of configurations. The lattice is updated sequentially. 
To reduce autocorrelations and accelerate
thermalization we include over-relaxation sweeps where the Metropolis
trial steps are taken deterministically in a way that approximately
inverts the action with respect to its minimum \cite{overrel} (exact
non-stochastic over-relaxation, as well as heat bath updating,
is not possible due to the non-linear
term in the action). For the Wilson lines, sweeps are performed by applying
$5$ random Metropolis hits and $2$ over-relaxed Metropolis
hits on each site before moving to the next site.
For the gauge links we mix $8$ random Metropolis hits
with $3$ over-relaxed hits per step. We treat the Wilson line $\boldl$
equivalent to the single time-like link of a $N_\tau=1$ gauge theory, for which
ref. \cite{Velytsky:2007gj} discussed the different treatments
of the time-like bondary conditions which are possible. We employ
the so-called \emph{time-plaquette-double-counting} scheme, where
the $U_{ij}$ possess staples in the positive as well as the negative
time-direction, which happen to give equal contributions. The result
is an additional factor two in front of the kinetic nearest
neighbour contribution to the action, when updating the spatial links $U_{ij}$.

All sample sizes quoted here are taken to be statistically independent
measurements.  We have estimated the integrated autocorrelation time
by using the
\emph{binning} method described in ref.~\cite{berg}. The
method consists of grouping fixed numbers of successive data points,
obtaining a new set of points which contains the average
values of each group and then comparing the variances of the old
and the new set. We found that on the order of $25$ configurations
must be discarded between measurements far from the phase
boundary and on the order of $400$ configurations
at the phase boundary on the largest ($N_s=24$) lattice (we refrain
from showing any figures). We found that
the thermal relaxation time within the Monte Carlo time series of
measurements is roughly of the same order not too close to the phase boundary. 

We have obtained error estimates for the 
numerical results presented in this work. The
errors of the phase boundary presented in
Fig. \ref{fig:Ph_Diagr} are given by the
finite resolution in $\beta$ and $m^2$ respectively.
The errorbars for the coefficients presented
in Figs. \ref{fig:coeff_1},\ref{fig:coeff_2},\ref{fig:k3coeff10},
\ref{fig:k3coeff20},\ref{fig:k3coeffm00},\ref{fig:k4coeff10}
and \ref{fig:coeff_e_f_g} result from the $\chi^2$ fit.
In Fig. \ref{fig:fst_snd_order2} we include the statistical
errorbar for the Polyakov expectation value. In all of these cases,
the errorbar is often smaller than the point size and thus cannot
be seen in the respective figure. The results
presented in Fig. \ref{fig:potmindiffs} are absolute numbers
and are presented without errorbars.
The potentials shown in Figs. \ref{fig:Ph_r}, \ref{fig:V_rho},
\ref{fig:blockspin_15_00}, \ref{fig:k3blockb10}, \ref{fig:k3blockb20}
and \ref{fig:fst_snd_order1} are obtained from a histogram, which is
obtained without statistical errorbars. However,
the errors can be estimated from the fluctuations which are visible
in the figures. We have obtained a data set of sufficient size, such that
fluctuations are strongly suppressed for a large region surrounding
the minimum of the potential. 

\section{Results}
\subsection{Phase diagram}

\begin{figure}[h!]
\includegraphics*[width=\linewidth]{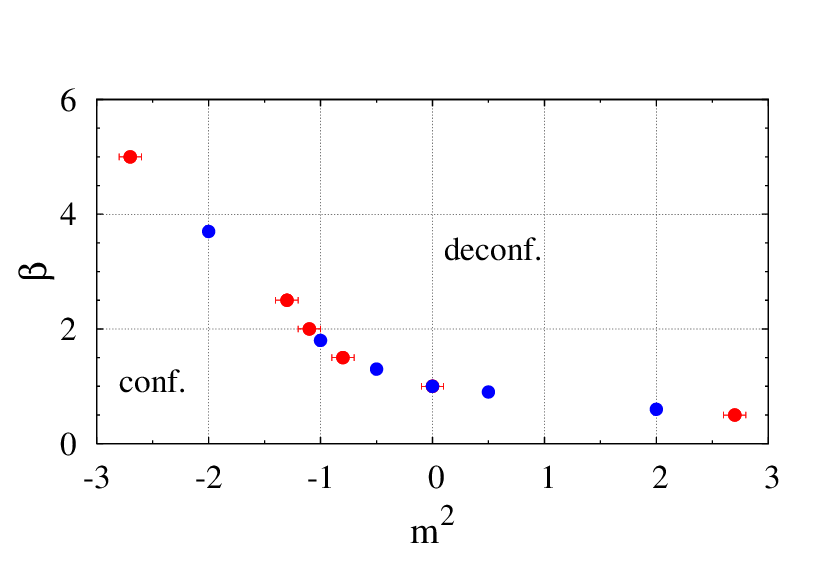}
\caption{The phase boundary in the $\beta-m^2$ plane. The Z(2)-broken
  phase corresponds to the upper-right region.\label{fig:Ph_Diagr}}
\end{figure}

To determine the phase diagram of the theory we measure the
expectation value of the Polyakov loop $\ell(\bm{x}) = \frac{1}{2}
\tr\boldl(\bm{x})$ and the inverse correlation length $m_\xi=1/\xi$ as
functions of the couplings $m^2$ and $\beta$. There is a line of phase
transitions in the $\beta-m^2$ plane separating the region where the
expectation value of the Polyakov loop vanishes (the Z(2) symmetric
phase at low $\beta$, $m^2$) from the Z(2) broken phase at large
$\beta$ or $m^2$ (see fig. \ref{fig:Ph_Diagr}).  The transition is of
second order since the inverse correlation length $m_\xi$ on the phase
boundary extrapolates to zero in the infinite volume limit \cite{DS}
as long as $\beta$ is not too large\footnote{The second order phase
  transition is further confirmed by the potentials described in later
  sections}.  It appears that the order of the phase transition
changes to first order at roughly $\beta > 3.0$.  This may signal a
breakdown of this model as an effective description of 4D Yang Mills
in the extreme weak coupling limit (for a discussion of this issue see
Appendix B).

Confinement is realized in distinct ways. At small $\beta$ and
vanishing $m^2$ we find $\langle\ell\rangle=0.0$ because the Wilson
line $\boldl(\bm{x})$ averages to zero from random fluctuations over
the whole group manifold. For large $\beta$ and negative $m^2$
(corresponding to the upper left region in fig.~\ref{fig:Ph_Diagr}), we
find a non-trivial Z(2) symmetric phase, where
the Wilson lines fluctuate only outside of the
group center ($\boldl(\bm{x})=i\tau_3$ or rotations thereof),
and the trace of $\boldl$ vanishes locally 
for each site\footnote{In other words, the distributions of the eigenvalues
  $\lambda_1$ and $\lambda_2$, which are gauge invariant, peak about
  $\pm1$.
  That distinct confined phases with different eigenvalue
  structure can also arise in 4d models of Polyakov loops
  coupled to gauge fields,
  was shown in refs.~\cite{Myers:2008zm}}.
Unlike for the model without gauge fields studied in ref. \cite{DS} 
however, there is no global alignment of the Wilson lines in the confined phase
and no sharp phase boundary separating the two types of confined
vacua. 

\subsection{Effective potential}
We are interested in the distribution of the eigenvalues $\lambda_1$,
$\lambda_2$ of the Wilson line $\boldl$ over the thermal ensemble of
field configurations at each site. Here, we shall focus on the
potential for the average of $\lambda_1$ and $\lambda_2$:
\bea
\rho(\bm{x}) = \frac{1}{2}\left|\lambda_1(\bm{x}) +
                             \lambda_2(\bm{x})\right|~~,
\eea
which for SU(2) is nothing but the absolute value of the Polyakov loop,
$\rho = \sqrt{\ell^2}$. From the probability distribution for $\rho$
we define an effective potential\footnote{Strictly speaking, what
is studied here is what by usual naming conventions is known as the
 ``constraint effective potential''.} via
\be \label{eq:Veff}
V_{\rm eff}(\rho) = -\log P(\rho)~~.
\ee
We normalize the probability density $P(\rho)$ over the interval [0,1]
which fixes the constant in the potential. This also factors out the 
volume dependence. The partition function can thus be written as
\bea
Z=\int d\rho\,e^{-N_s^3\,V_{\rm eff}(\rho)}
~.\label{eq:Veff_part}
\eea
In our definition, therefore, $V_{\rm eff}(\rho)$ is dimensionless
as it absorbs the volume $a^3$ of a lattice cell and, implicitly, a
factor of $1/T$ (because we employ a 3d theory).

Below, we shall show that in the weak-coupling regime (large $\beta$,
small $m^2$) a non-analytic contribution $\sim \sqrt{\ell^2}$ to the
effective potential arises dynamically. It is distinct from the
Vandermonde potential $V_{\rm Vdm} = -\frac{1}{2}\log(1-\ell^2)$
generated by the SU(2) group measure, and from the "bare" potential $\sim
-m^2 \ell^2$ which is included in the action~(\ref{Slatt}). We
find that for a broad range of the couplings $\beta$ and $m^2$ the
potential has the form
\be \label{eq:Vrho}
V_{\rm fit}(\rho) =  -\frac{1}{2}\log(1-\rho^2) \, +\,  a-b \rho+c \rho^2~~.
\ee
Note that the term proportional to $\rho\equiv\sqrt{\ell^2}$ of course
does not break the Z(2) symmetry explicitly, and is not to be confused
with a Z(2) background field $\sim \, -h\ell$ corresponding to (heavy)
dynamical quarks in the fundamental representation (note also
that the coefficient $a$ is just a normalization and is not
of any physical significance).

All measurements presented here were performed on a $N_s=24$ cubic
lattice. However, using smaller lattices we have checked that the
coefficients $a$, $b$, $c$ from~(\ref{eq:Vrho}) do not change much with
volume if $N_s \geq 12$. $N_s=24$ appears to be a good approximation
to the infinite volume limit where, as indicated in eq.~(\ref{eq:Veff_part}),
one expects the potential to be volume independent
(this holds even close to the phase boundary
where correlation lengths diverge). 5000 independent
lattice configurations were generated for each combination of 
$\beta, m^2$.

\begin{figure}[h!]
\includegraphics*[width=\linewidth]{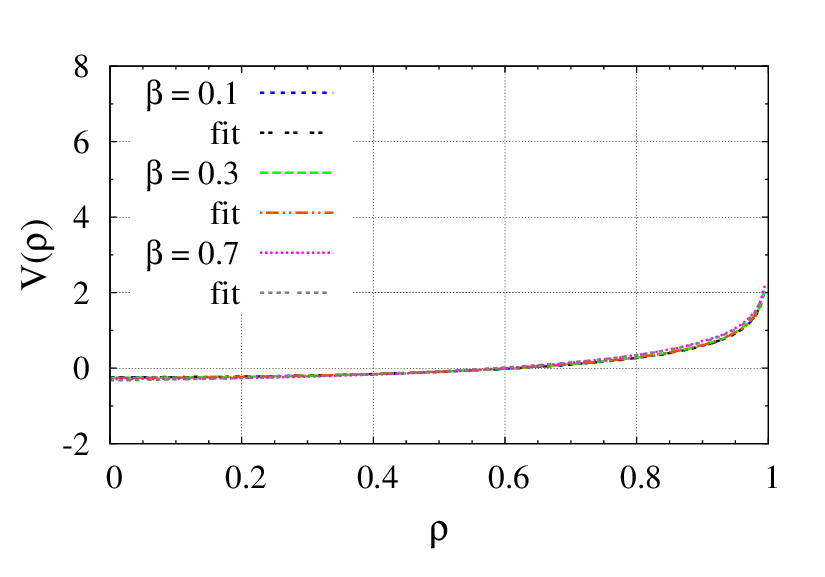}
\includegraphics*[width=\linewidth]{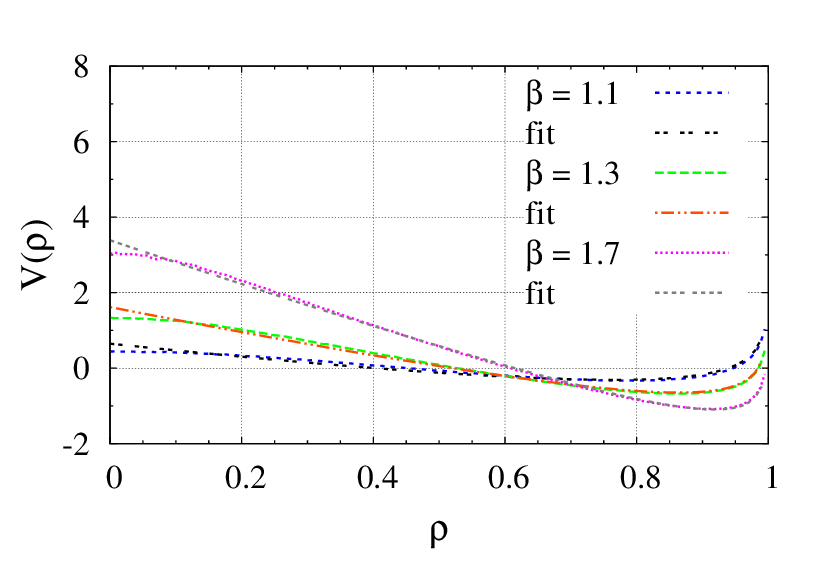}
\caption{Effective potential for $\rho= \sqrt{\ell^2}$ at $m^2=0$ for
  various values
  of $\beta$ fitted with the {\em ansatz}
  (\ref{eq:Vrho}). For $\beta<\beta_C$ (top) one sees only the
  group integration measure. Above $\beta_C$ spontaneous breaking of 
  Z(2) is evident. \label{fig:Ph_r}}
\end{figure}

\begin{figure}[h!]
\includegraphics*[width=\linewidth]{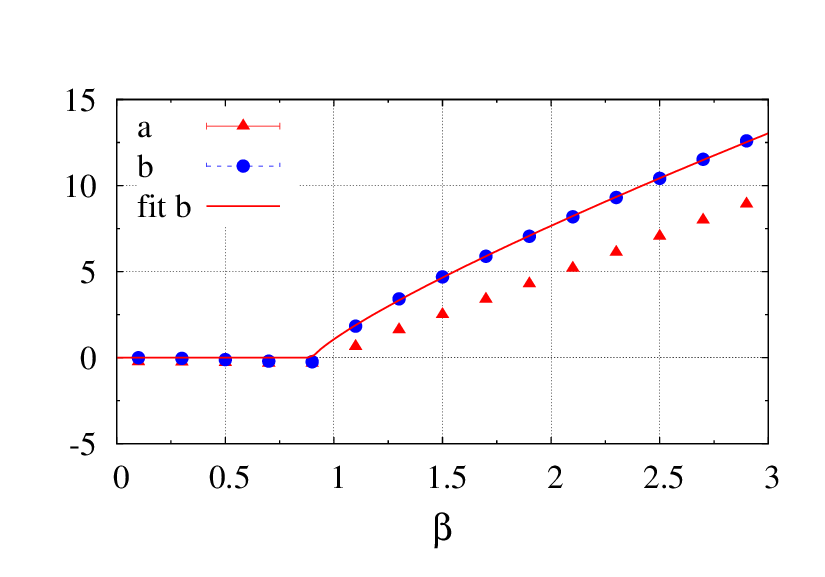}
\caption{Coefficients of the effective potential $V(\rho)$ for $m^2=0$ as a
	function of $\beta$. Below $\beta_C$ there is no deviation
	from the Vandermonde potential. Above $\beta_C$ one
        sees a non-analytic contribution. (A
        discussion of the fit-curve is found in Appendix
        A.) \label{fig:coeff_1}} 
\end{figure}

We first consider the case without potential, $m^2=0$. The phase
transition occurs at
$\beta_C \approx 0.9$ for this case.  Below the phase
transition point we find that
the potential defined via eq.~(\ref{eq:Veff}) coincides with the
Vandermonde potential $V_{\rm Vdm}$, hence $a=b=c=0$ within numerical
precision. This is shown in fig.~\ref{fig:Ph_r}.
On the other hand, for $\beta>\beta_C$ both coefficients $a$ and $b$
from eq.~(\ref{eq:Vrho}) turn non-zero. The quadratic coefficient $c$
does not appear since $m^2=0$. The coefficient $b$ appears to arise
purely from the dynamics of fluctuations. Its behavior is shown in
fig.~\ref{fig:coeff_1}.

\begin{figure}[h!]
\includegraphics*[width=\linewidth]{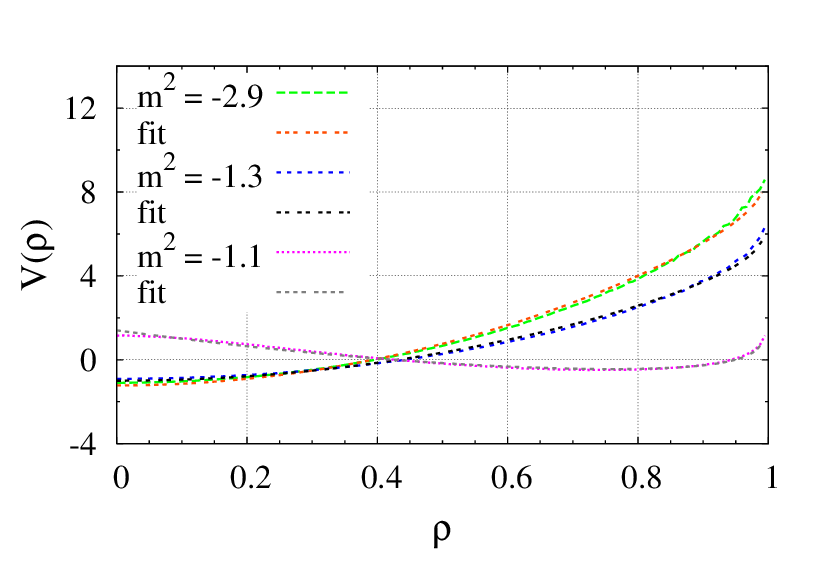}
\includegraphics*[width=\linewidth]{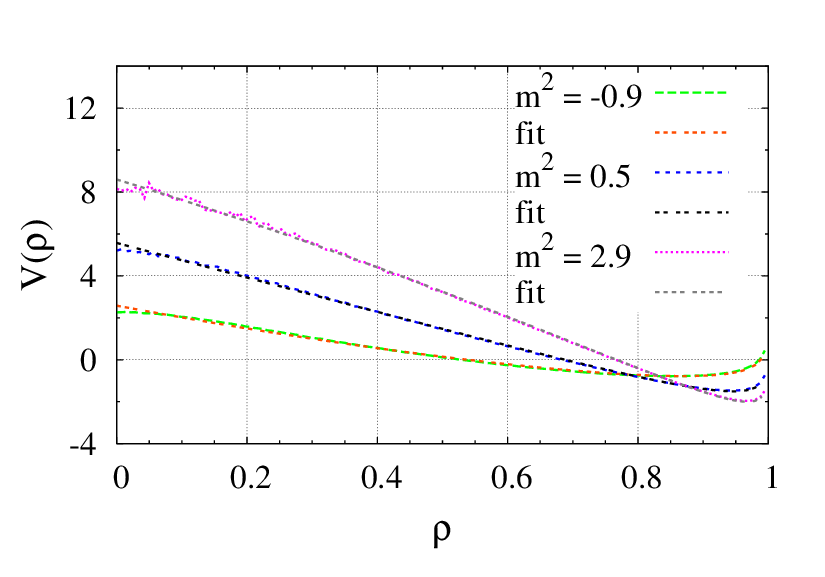}
\caption{Effective potential for $\rho$ at $\beta=2.0$ for various values
  of $m^2$ above and below the phase transition, fitted with the {\em ansatz}
  (\ref{eq:Vrho}). \label{fig:V_rho}}
\end{figure}

The {\em ansatz}~(\ref{eq:Vrho}) also works for $m^2\neq 0$ when $\beta$ is not
too large. We have
confirmed this for a broad range of $m^2$ for several fixed values of
$\beta$. Explicit results for $\beta=2.0$ are shown in fig.~\ref{fig:V_rho}.
The coefficient $c$ of the quadratic term appears to depend linearly
on the square mass from the bare potential, $c \sim -m^2$, 
except exactly at the phase transition where it changes discontinously.

However, we again observe that above the phase
transition point the dynamics generates a non-analytic contribution to
the potential for the Polyakov loop proportional to $\rho\equiv
\sqrt{\ell^2}$. For illustration, we show the $m^2$ dependence of
$a,b,c$ for $\beta=1.0$ and $\beta=2.0$ in fig.~\ref{fig:coeff_2}. A
detailed discussion of the dependence of $a,b,c$ on $\beta$ and $m^2$
is given in appendix A.  The main point here is that the Vandermonde
contribution to the effective potential does not depend on $\beta$ and
$m^2$, and that the effective potential obtained from a histogram of
$\rho({\bf x})$ at each lattice site is different from a
Landau-Ginzburg type mean-field theory for the Polyakov loop. In the
next section we shall see that when the field configurations are
``cooled'' to remove short wavelength fluctuations, that in fact one
does obtain a potential that resembles mean field theory, but with a
coefficient multiplying the Vandermonde potential which depends on
$\beta$ and $m^2$.
\begin{figure}[h!]
\includegraphics*[width=\linewidth]{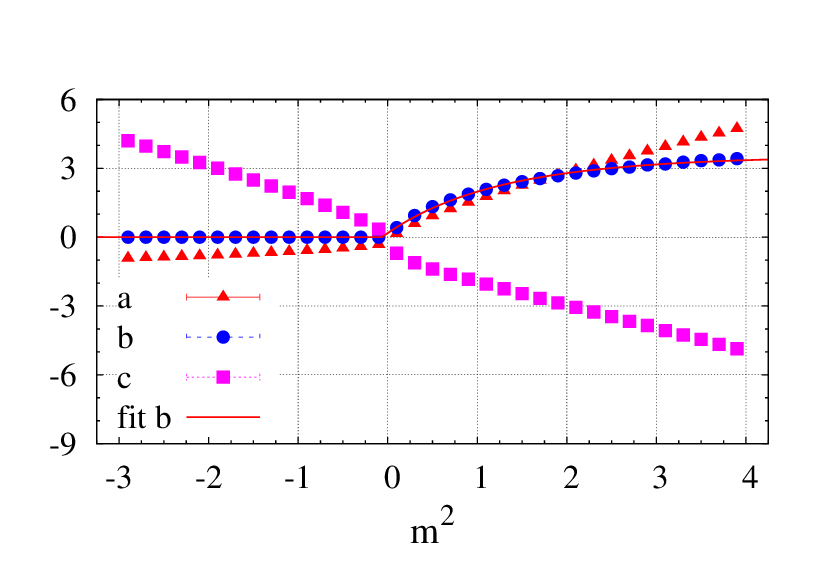}
\includegraphics*[width=\linewidth]{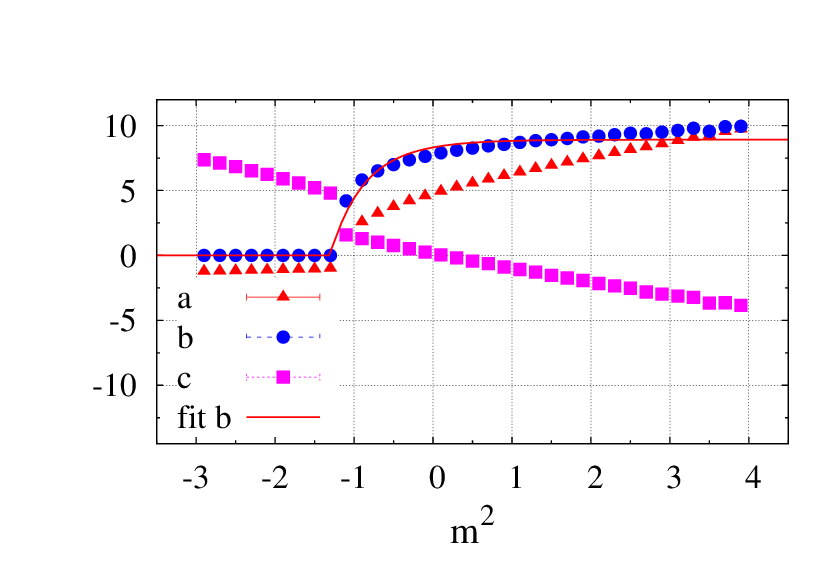}
\caption{Coefficients of $V(\rho)$ for $\beta=1.0$ (top) and
  $\beta=2.0$ (bottom) as functions of $m^2$. The non-analytic
  contribution sets in at the phase transition.\label{fig:coeff_2}}
\end{figure}
\subsection{Effective potential for block spins}

\begin{figure}[h!]
\includegraphics*[width=\linewidth]{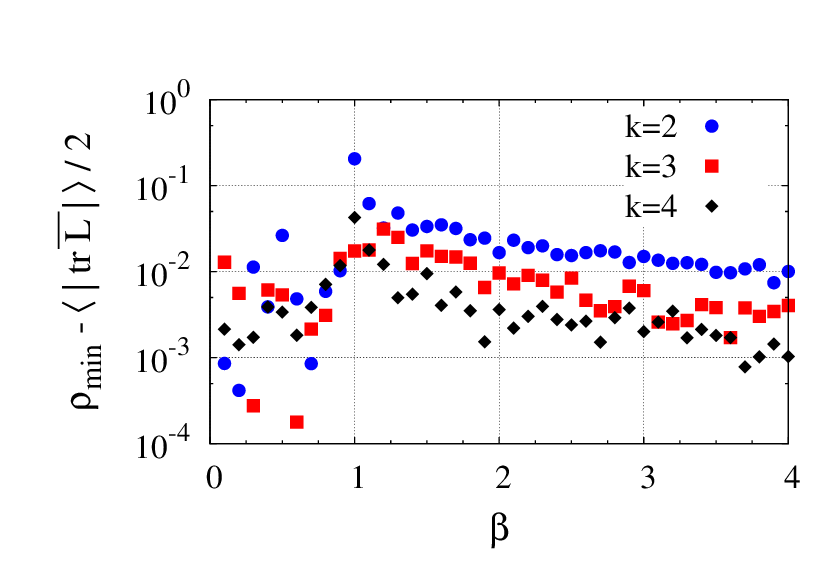}
\includegraphics*[width=\linewidth]{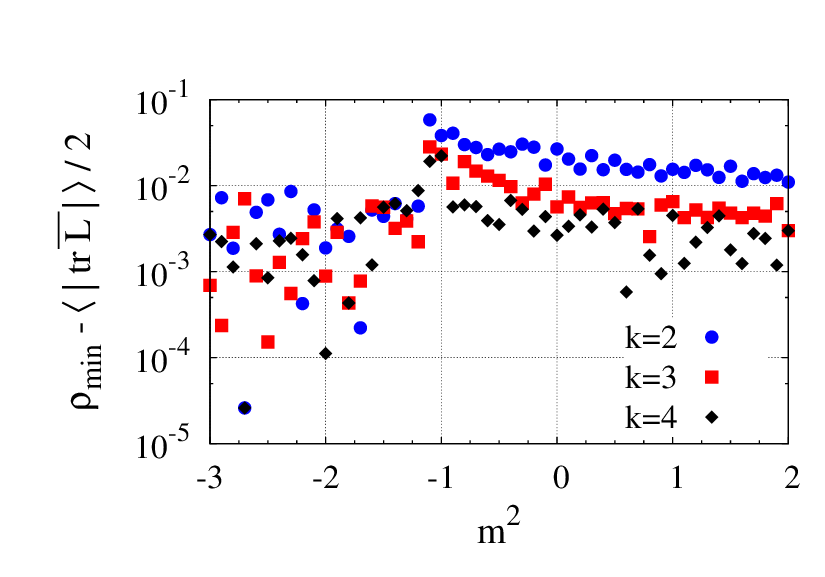}
\caption{The difference of Polyakov loop expectation value
and minimum of effective potential after ``cooling'' for
$k=2,3,4$ at $m^2=0.0$ as a function of $\beta$ (top) and
at $\beta=2.0$ (bottom) as a function of $m^2$.
The peaks correspond to the phase transition point.
\label{fig:potmindiffs}}
\end{figure}

\begin{figure}[h!]
\includegraphics*[width=\linewidth]{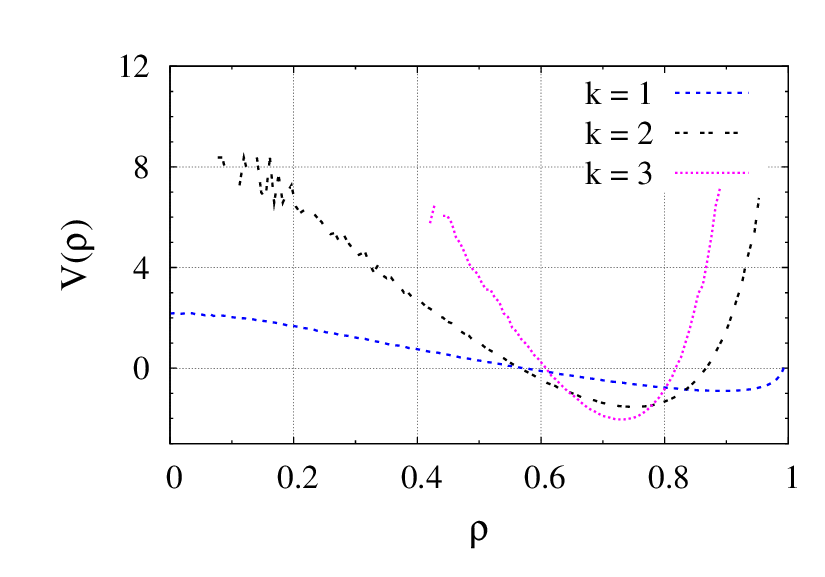}
\caption{Blockspin averages for $\beta=1.5$, $m^2=0.0$. As $k$ increases
the potential peaks more sharply about the
expectation value. \label{fig:blockspin_15_00}}
\end{figure}

To obtain a potential for the {\em long wavelength modes} we average
the Polyakov loop field over small cubes of side-length $k$ before
histogramming. This eliminates the short wavelength
spatial field modes.  We calculate \emph{blockspin averages} defined
as 
\bea \bar{\ell}_i^{(k)}=\frac{1}{k^3}\sum_{\vec{n}} \frac{1}{2} \tr
\boldl(\vec{i}+\vec{n})~,\\ \vec{n}=(0,0,0) \ldots (k,k,k)~.\nonumber
\eea 
We have investigated the cases $k=2,3,4$. Blockspins were measured on
2500 independent configurations for each combination of $\beta,
m^2$. As one expects from the central limit theorem (at least
in the region far from the phase boundary where screening
masses are large and individual lattice sites fluctuate independently), with increasing
$k$ the potential becomes more symmetric and peaked around the actual
expectation value (see fig.~\ref{fig:blockspin_15_00}). In what
follows, we take the configurations for
$k=3$ as a good approximation for the long distance sector. To see this
consider fig.~\ref{fig:potmindiffs}.
At $k=3$ the minimum of the fitted effective potential from
eq.~(\ref{eq:Vrho_meanfield}) differs at most by $\simeq 0.03$ from
the numerical result for the Polyakov loop expectation value
$\langle\ell\rangle$. This maximal deviation occurs exactly at the
phase transition point. Here, within our numerical precision, $k=4$
does not do significantly better. Away from the phase transition,
$k=3$ differs less than $\simeq 0.01$ from the numerical value of
$\langle\ell\rangle$. $k=4$ does slightly better but reduces our
statistics significantly. All results presented below appear to be
stable when going from $k=3$ to $k=4$ (we will discuss an explicit
example below). We refrain from discussing $k=2$ in detail here as it
appears that contributions from the short range fluctuation have not
yet been completely eliminated.

\begin{figure}[h!]
\includegraphics*[width=\linewidth]{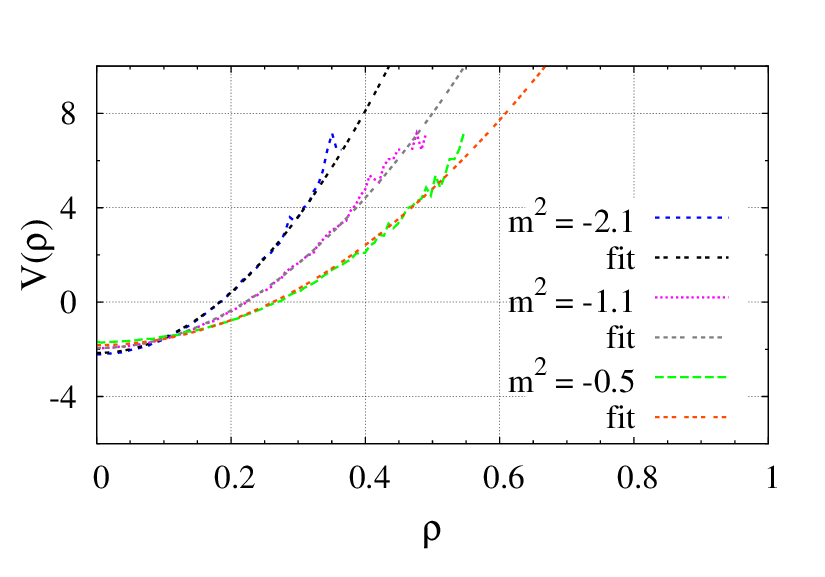}
\includegraphics*[width=\linewidth]{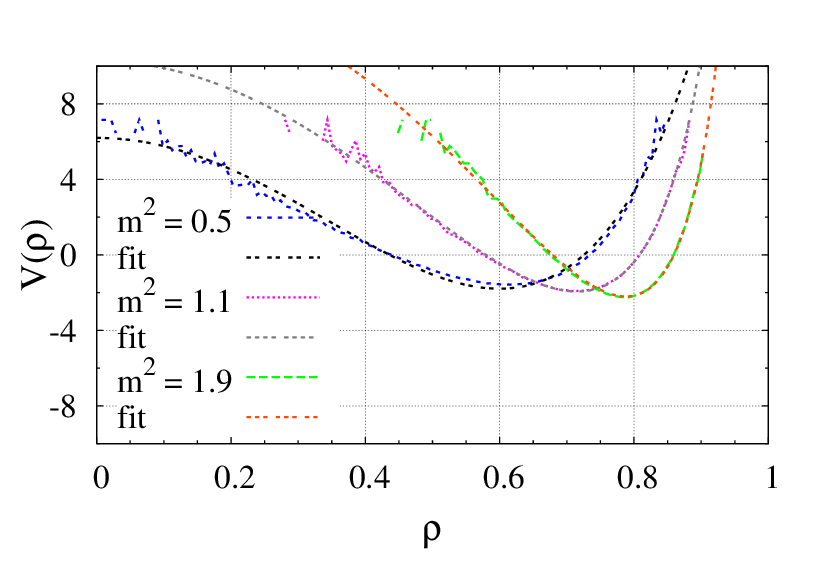}
\caption{The potential for blocks of side length $k=3$ fit with
ansatz (\ref{eq:Vrho_meanfield}) for $\beta=1.0$
in confined (top) and deconfined (bottom) phases.
\label{fig:k3blockb10}}
\end{figure}

\begin{figure}[h!]
\includegraphics*[width=\linewidth]{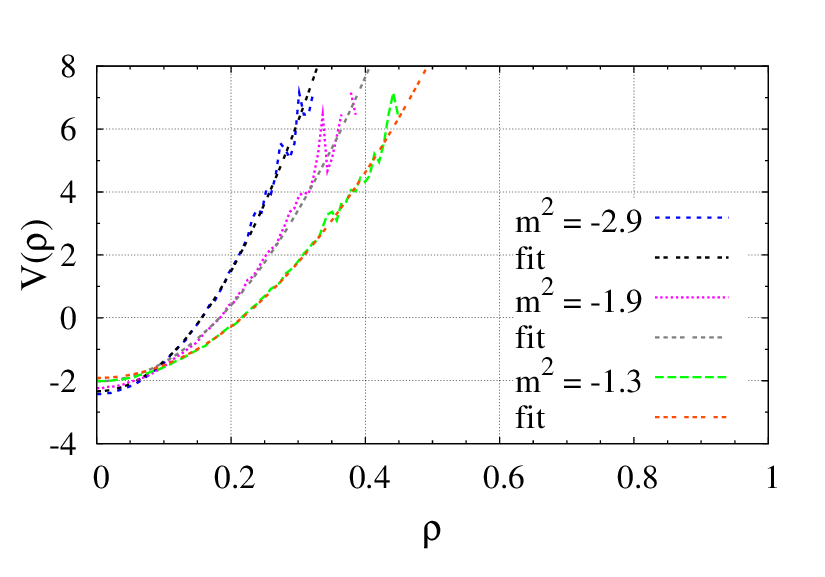}
\includegraphics*[width=\linewidth]{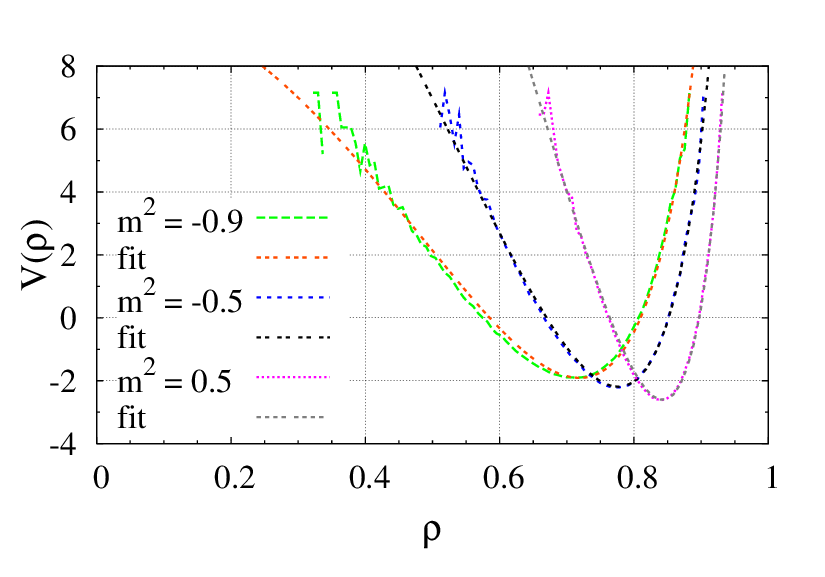}
\caption{The potential for blocks of side length $k=3$ fit with
ansatz (\ref{eq:Vrho_meanfield}) for $\beta=2.0$
in confined (top) and deconfined (bottom) phases.
\label{fig:k3blockb20}}
\end{figure}

The {\em ansatz}~(\ref{eq:Vrho}) is no longer applicable for the long
wavelength modes. The blockspin averaging appears to suppress the
non-analytic term in the potential in most parts of the phase diagram
(some possible exceptions are discussed below). It appears that over a
broad range of $\beta$ and $m^2$ the potential can be described fairly
well by a form analogous to mean field models, with quadratic and
quartic terms~\cite{Svetitsky:1982gs}. However, for a good fit to
the extracted potential one needs to include an additional parameter
$d_0$ into the fit function which varies with $\beta$ and $m^2$, and
which multiplies the Vandermonde potential. So, finally, the form
which we use to model our data is
\be \label{eq:Vrho_meanfield}
V(\rho) =  -d_0\frac{1}{2}\log(1-\rho^2) \, +\,  d_1+d_2 \rho^2+d_4 \rho^4~~.
\ee
Here, a linear term is not included. We also point out that the
quartic term arises from the dynamics of fluctuations as such a
contribution is not included in the ``bare'' action~(\ref{Slatt}).
Explicit results for the
numerical potential and for the fit via eq.~(\ref{eq:Vrho_meanfield}),
for different values of $m^2$ in the confined and deconfined phases
and fixed $\beta=2.0~/~1.0$, are shown in
figs.~\ref{fig:k3blockb10},\ref{fig:k3blockb20}. Below the phase
transition one can set $d_0=d_4=0$ and fit the potential with a simple
quadratic form.
This is not surprising since for $\langle\ell\rangle=0.0$ the
potential is essentially parabolic and there is little sensitivity
to higher powers of $\ell$.

\begin{figure}[h!]
\includegraphics*[width=\linewidth]{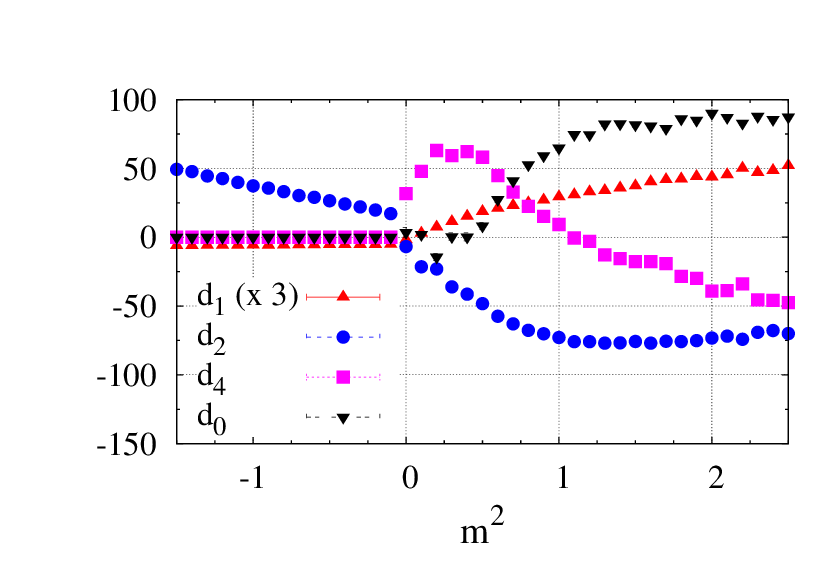}
\caption{The coefficients in eq.~(\ref{eq:Vrho_meanfield}) for $\beta=1.0$.
The Vandermonde term appears to gradually rise to its asymptotic
value. \label{fig:k3coeff10}}
\end{figure}
\begin{figure}[h!]
\includegraphics*[width=\linewidth]{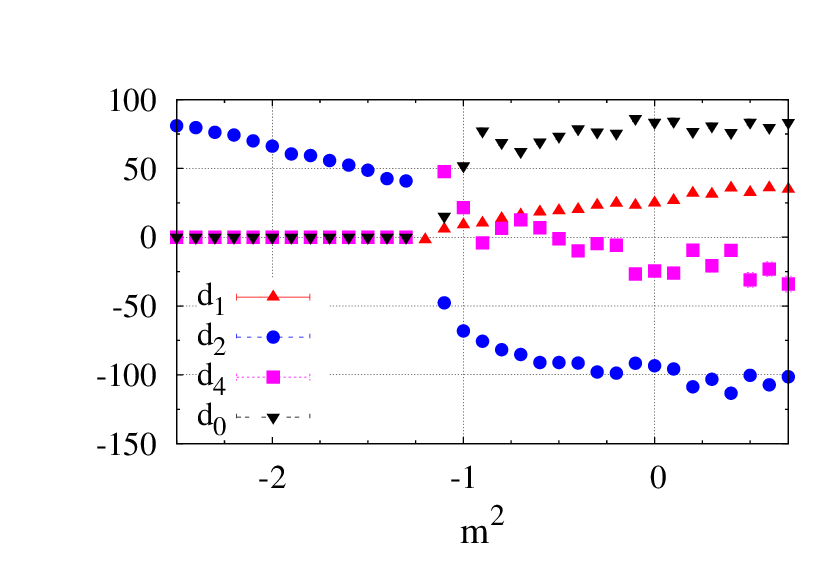}
\caption{The coefficients of eq. (\ref{eq:Vrho_meanfield}) $\beta=2.0$.
The region with vanishing Vandermonde term appears to shrink when
going from $\beta=1.0$ to $\beta=2.0$. Compare to
fig.~\ref{fig:k3coeff10}
\label{fig:k3coeff20}}
\end{figure}
\begin{figure}[h!]
\includegraphics*[width=\linewidth]{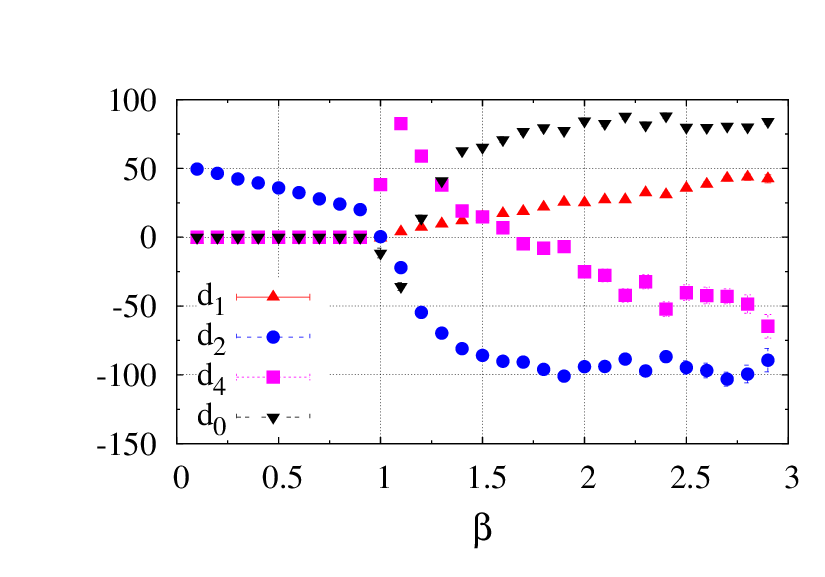}
\caption{A region with vanishing Vandermonde term is visible
for $m^2=0.0$ and $\beta > \beta_C$. 
\label{fig:k3coeffm00}}
\end{figure}
We show how the fit parameters from
eq.~(\ref{eq:Vrho_meanfield}) depend on $\beta$ and $m^2$.
Figs.~\ref{fig:k3coeff10} and \ref{fig:k3coeffm00} correspond to
$\beta=1.0$ with variable $m^2$ and $m^2=0.0$ with variable $\beta$,
respectively. One observes that right above the phase transition, the
potential is a sum of quadratic and quartic terms while the
Vandermonde contribution vanishes (in the figures, $d_0$ appears to
fluctuate somewhat around zero. We have checked, however, that the
result is consistent with setting $d_0=0$ by hand). At higher $m^2$ or
$\beta$ respectively, the coefficient $d_0$ increases gradually and
saturates at about $d_0\approx 80$ for large values of $\beta$ or
$m^2$, while the quartic coefficient becomes negative. We have further
checked that fixing $d_0$ to its asymptotic value ($d_0\approx80$)
gives a less accurate fit and increases $\chi^2$ per degree of freedom
in the region closely above the phase transition roughly by a factor
of two. This indicates that the gradual increase of $d_0$ is a real
dynamical effect and not just an artifact generated by a lack of
sensitivity to the Vandermonde potential when $\langle\rho\rangle$ is
much smaller than 1.  Fig.~\ref{fig:k3coeff20} shows that the region
right above the phase transition, where the Vandermonde vanishes,
appears to shrink when going deeper into the weak coupling limit
(larger $\beta$).

The question remains, how this suppression of $d_0$ just above the
phase transition, at moderately weak coupling, comes about. We have
therefore attempted to model the potential right above the transition
with a different function, assuming a fixed Vandermonde potential
term, equal to the asymptotic value, but also allowing additional
terms. We find that it is possible, within our numerical accuracy, to
trade the suppression of the Vandermonde for another term linear in
$\rho$. The function
\bea 
&V(\rho)=  -d_0\frac{1}{2}\log(1-\rho^2)  +\, d_1
 +\,d_0' \rho +d_2 \rho^2+d_4 \rho^4,~\nonumber\\
& ~~~\textrm{with} ~~~ d_0\equiv 80~,\label{eq:Vrho_meanfield2}
\eea
reproduces the behavior of the effective potential
around $\rho\approx 0.0$ even slightly better than
eq.~(\ref{eq:Vrho_meanfield}), with a negative coefficient $d_0'$ in the
region right above the phase transition. However, the
improvement in $\chi^2$ is below the percent level
and the function (\ref{eq:Vrho_meanfield2}) fails completely
at large $\beta$ or $m^2$ (by generating a potential that is not
bounded from below).

\begin{figure}[h!]
\includegraphics*[width=\linewidth]{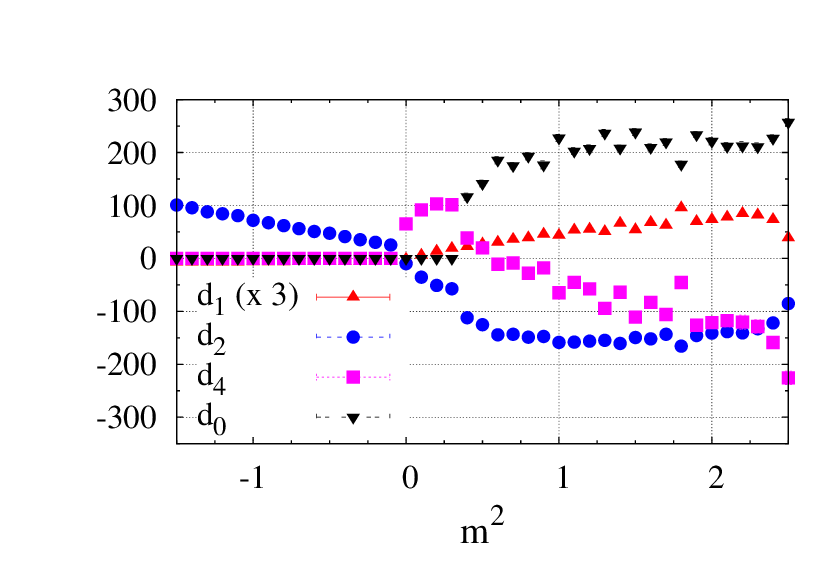}
\caption{The coefficients of eq.~(\ref{eq:Vrho_meanfield}) for $\beta=1.0$
at $k=4$. Compare to fig.~\ref{fig:k3coeff10}. For $m^2\leq 0.3$ we
set $d_0=0$ by hand (since in this region it yields no improvement
of the goodness of the fit). \label{fig:k4coeff10}}
\end{figure}

The function (\ref{eq:Vrho_meanfield2}) may suggest that the
suppression of the Vandermonde could be an artifact due to incomplete
cooling of short-distance fluctuations. However, we have investigated
the cases [$\beta=1.0$/variable $m^2$] and [$m^2=0.0$/variable
  $\beta$] also for $k=4$ and obtained similar results, up to an
overall scaling factor for all coefficients in the
potential\footnote{A rescaling of the coefficients in the potential is
  expected because coarse-graining over $k$ lattice sites
  effectively corresponds to a simulation with different lattice
  spacing. A quantitative comparison of the coefficients of the $k=3$
  potential to those of the $k=4$ potential would also require a
  rescaling of $\beta$.}. We show the result for $\beta=1.0$ in
fig.~\ref{fig:k4coeff10}. Compared to fig.~\ref{fig:k3coeff10} we only
observe a slight suppression of the quartic coefficient right above
the phase transition.
\section{Summary and Discussion}

We have performed simulations of an effective theory of Wilson lines
coupled to gauge fields in three dimensions which respects the center
symmetry of the four-dimensional SU(2) Yang-Mills theory. After
mapping the phase diagram, we have investigated the effective
potential for the average of the eigenvalues of the SU(2) Wilson line,
which is equal to the absolute value of the Polyakov loop.  We found
that a form containing non-analytic contributions can describe the
extracted potential. This non-analytic term was not present in the
action, and therefore must arise from the dynamics.

We extracted a similar effective potential also for the long
wavelength modes of the Polyakov loop and found that this can be
described by a mean-field type potential with quadratic and quartic
terms plus an effective Vandermonde potential which depends on the
couplings. Just above the phase boundary, in the deconfined phase, the
effective Vandermonde potential contributes little. Deeper into the
deconfined phase its coefficient increases and eventually appears to
approach a constant.

Our simulations may provide useful insight into the structure of
mean-field type models for the deconfining phase transition. For
example, so-called ``Polyakov-NJL'' models have recently been studied
extensively. Such models attempt to describe QCD thermodynamics over a
range of quark masses, from the pure-gauge limit to physical QCD; they
require an {\em ansatz} for the effective potential for the Polyakov
loop. For example, in early works on this
subject~\cite{Fukushima:2003fw} a quadratic potential for $\ell$ has
been used, plus a Vandermonde contribution (per lattice site) which is
constant and temperature independent. Our results appear to indicate,
however, that if a standard potential with terms $\sim\ell^2$ and
$\sim\ell^4$ (plus cubic Z(3) invariants for the case of three colors)
is used, that a temperature dependent Vandermonde contribution should
also be allowed for (this is already applied in some recent works,
see e.g. ref. \cite{Hell:2009by} and references therein).
The present theory does not include fermions and so it
is unfortunately not possible to address their effect. Nevertheless,
PNJL models generally also include a pure loop potential, which
describes the deconfining phase transition in the limit of inifinitely
heavy fermions.

Quantitative results for the physical case of three colors will of course
differ, and indeed the order of the
deconfining phase transition is different. Nevertheless, it appears
hard to imagine that qualitative aspects, for example a temperature
dependence of the effective Vandermonde term in the fitted effective
potential, would be absent.

\section*{Acknowledgements}
I am indebted to Adrian Dumitru, Rob Pisarski and Stefan Schramm for
many helpful
discussions. The numerical simulations presented
here were performed at the Center for Scientific Computing (CSC) at
Frankfurt University. Our code is based in part on the MILC
collaboration's public lattice gauge theory code, see 
http://physics.utah.edu/\~{}detar/milc.html.

\appendix
\section{Coefficients in the effective potential for all modes}
In this appendix we discuss the behavior of the non-analytic term
in the potential defined in eq.~(\ref{eq:Vrho})
as a function of $\beta$ and $m^2$.

The case $m^2=0.0$ is rather simple. The nearly
linear behavior of $b(\beta)$ above $\beta_C$, seen
in Fig.~\ref{fig:coeff_1} suggests
an {\em ansatz} of the form
\be
b(\beta) = b_0 \, (\beta-\beta_C)^r \, \theta(\beta-\beta_C)~.
\label{eq:coeff_1}
\ee
Indeed, we find that with $\beta_C=0.9$, a good fit of $b(\beta)$ is
possible, resulting in
$b_0 =7.1(1)$ and $r=0.82(4)$. This fit corresponds to the solid
line in fig.~\ref{fig:coeff_1}.

For $m^2 \neq 0.0$ the situation is more involved. 
As a heuristic analogy to the 2D Ising model we try the {\em ansatz}
\begin{multline}b(\beta,m^2)=\tilde{b}(\beta)\,
\theta(m^2+\tilde{m}^2(\beta))\\
  \times  \left\{1-\left[ \sinh \left( 2\left[ g(\beta)\,
 \left(m^2+\tilde{m}^2(\beta)\right)+\tilde{\beta}_C \right] \right)
  \right]^{-5} \right\}\\
  \label{eq:linear} 
\end{multline}
with $\tilde{\beta}_C=\log(1+\sqrt{2})$ (see fig.~\ref{fig:coeff_2}). This is
similar to the magnetization in the 2D Ising model~\cite{Fiore:1998uk}, which
is given by 
\be
M(\beta)=\theta(\beta-\tilde{\beta}_C)\, (1-[\sinh(2\beta J
)]^{-4})^{\frac{1}{8}}~.\label{eq:ising}
\ee
We find that~(\ref{eq:linear}) works reasonably well for $\beta \leq
3.0$ (see the fit-curves in fig.~\ref{fig:coeff_2} for specific
examples). Note however that the exponents in eq. (\ref{eq:linear})
and eq. (\ref{eq:ising}) differ and that the \emph{ansatz} (\ref{eq:linear})
does not imply any deeper connection from universality arguments.
Once the $\beta$ dependences of the parameters
$\tilde{m}^2$, $g$ and $\tilde{b}$ of eq. (\ref{eq:linear}) have
been obtained it also describes $b(\beta,m^2=0)$, although less accurately
than eq. (\ref{eq:coeff_1}) since separate data sets (with fixed $\beta$
and variable $m^2$)
containing mostly measurements far from $m^2=0.0$ were used to
obtain $\tilde{m}^2(\beta)$, $g(\beta)$ and $\tilde{b}(\beta)$.
There is no straight-forward way to obtain eq. (\ref{eq:coeff_1})
analytically from eq. (\ref{eq:linear}) and fitting (\ref{eq:linear})
directly to $b(\beta, m^2=0)$ is not feasible due to the large number
of additional parameters that enter with the $\beta$ dependence of
$\tilde{m}^2$,$g$ and $\tilde{b}$.

We have included the constant $\tilde{\beta}_C$ in
eq.~(\ref{eq:linear}) because it corresponds to the critical point of
the Ising model. Our model of course deconfines at a different value
of $\beta$. Nevertheless, an {\em ansatz} such as eq.~(\ref{eq:ising})
implicitly assigns the number $\tilde{\beta}_C$ a special meaning and
we therefore include it into our ansatz also in order to ``filter
out'' its effect. Isolating $\tilde{\beta}_C$ in such a way simplifies
the resulting dependence of the fit parameters on $\beta$ greatly.

\begin{figure}[h]
\includegraphics*[width=\linewidth]{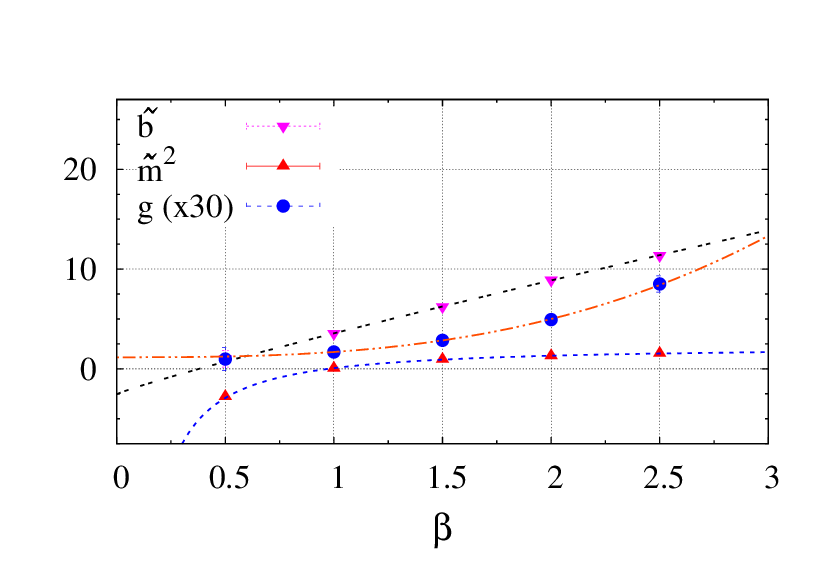}
\caption{$\beta$ dependence of various coefficients parameterizing
  $b(\beta,m^2)$ with their corresponding fit curves.
   $g(\beta)$ is scaled up by a factor of 30.
   \label{fig:coeff_e_f_g}}
\end{figure}

The coefficients introduced in eq.~(\ref{eq:linear}) act as follows:
$\tilde{m}^2$ corresponds to a shift along the horizontal
axis. $\tilde{b}$ is a scaling factor and $g$ is a modification of the
coupling strength.  The $\beta$ dependence of these coefficients can
be described by power laws
\bea
\tilde{m}^2(\beta)&=& m_0^2 + m_0^{\prime\,2} \beta^{w}~,\\
\tilde{b}(\beta)&=& b_0^{\prime}+ b_0^{\prime\prime}\beta^v~, \\
g(\beta)&=&g_0+g_0^{\prime}\beta^u~,
\eea
with
\bea
m_0^2&=&2.2(1)~,~m_0^{\prime\,2}=-2.1(1)~,~w=-1.2(1)~,\\
b_0^{\prime}&=&-2.5(4)~,~b_0^{\prime\prime}=6.0(4),~v=0.90(4)~,\\
g_0&=&0.038(1)~,g_0^{\prime}=0.017(1) ~,~u=2.8(1) ~.
\eea
The $\beta$ dependence of $\tilde{m}^2$, $\tilde{b}$ and $g$ is shown in 
fig.~\ref{fig:coeff_e_f_g}.\\

\section{First-order transition in extreme weak coupling limit}

\begin{figure}[h]
\includegraphics*[width=\linewidth]{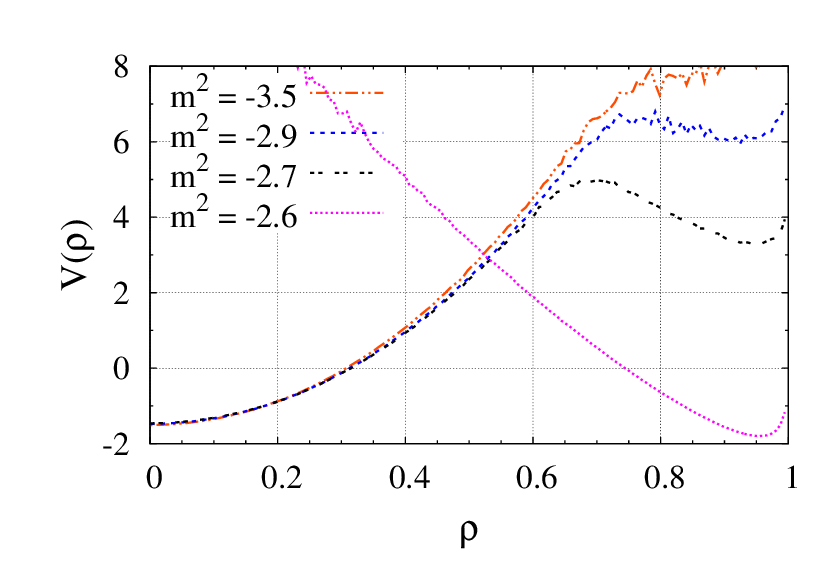}
\caption{Effective potential for $\beta=5.0$ at values of $m^2$
  slightly above and slightly below the phase transition. A first order
  phase transition is apparent since one observes two distinct
  minima.}
\label{fig:fst_snd_order1}  
\end{figure}

\begin{figure}[h!]
\includegraphics*[width=\linewidth]{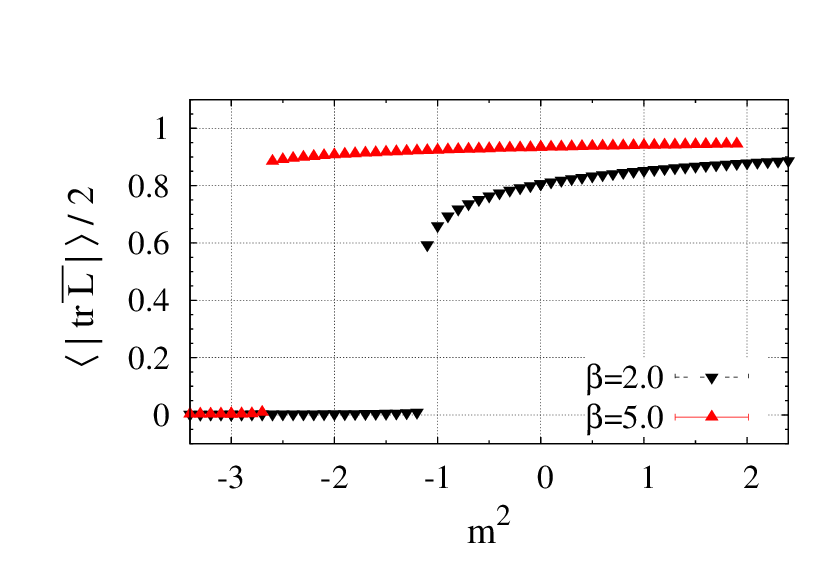}
\caption{Polyakov Loop expectation value for $\beta=2.0$ and $\beta=5.0$
measured on $N_s=24$. The transition becomes very sharp
for large $\beta$.} \label{fig:fst_snd_order2}
\end{figure}

At very large $\beta$ the transition becomes first order. To see this,
consider figs.~\ref{fig:fst_snd_order1} and \ref{fig:fst_snd_order2}.
For $\beta=5.0$ the effective potential develops two distinct minima
in the vicinity of the phase transition point. Correspondingly, the Polyakov
loop expectation value appears to be discontinuous.
This behavior is in sharp contrast to the case $\beta=2.0$ (shown in
fig.~\ref{fig:V_rho}), where one can see a single minimum
moving continuously with $m^2$ (within the resolution) from $\rho\approx0$ to
$\rho\approx1$.

The first order transition indicates that our effective theory
cannot describe 4D Yang Mills when $\beta$ is too large. 
This is not problematic however, since the coupling constant $\beta$
in the effective theory is unrelated to the 4D case. When matching the theories,
any value of $\beta$ in the 4D SU(2) Yang Mills theory close to the
second order phase transition will correspond to a combination of coupling
constants in the effective theory in the vicinity of the second order phase boundary. 
While it may be necessary to include higher powers of the Polyakov loop 
and additional coupling constants to the action~(\ref{Slatt}) to make a precise
matching possible (there is also no ad hoc justification for the assumption
that the coupling constants of the kinetic energy term and of the 3D Wilson action of the
spatial gauge fields must be the same), the presence of a first order phase boundary in the parameter
space of the effective theory is itsself of no relevance to the matching procedure.

\end{document}